\documentstyle[prl,aps,psfig]{revtex}
\begin{document}
\draft
\title{Ergodicity properties of energy conserving single spin flip
dynamics in the $XY$ model}
\author{ Abhishek Dhar }
\address{Theoretical Physics Group, Tata Institute of Fundamental Research,
Homi Bhabha Road, Bombay 400005, India. \\
e-mail : abhi@theory.tifr.res.in}
\date{\today}

\maketitle
\widetext
\begin{abstract}
A single spin flip stochastic energy conserving dynamics for the $XY$
model is considered. We study the ergodicity properties of the
dynamics. It is shown that phase space trajectories densely fill the
geometrically connected parts of the energy surface. We also show that   
while the dynamics is discrete and the
phase point jumps around, it cannot make transitions between closed
disconnected parts of the energy surface. 
Thus the number of distinct sectors depends on the number of
geometrically disconnected parts of the energy surface. Information on
the connectivity of the surfaces is obtained by studying the
critical points of the energy function.
We study in detail the case of two spins and find that the number of
sectors can be either one or two, depending on the external fields and
the energy. 
For a periodic lattice in $d$ dimensions, we find regions in phase
space where the dynamics is non-ergodic and obtain a lower bound on
the number of disconnected sectors. We provide some numerical evidence
which suggests that such regions might be of small measure so that the
dynamics is effectively ergodic.
\end{abstract}



\section{Introduction }

A much studied problem in non-equilibrium statistical mechanics is
the problem of heat conduction. Here one is interested in obtaining
the steady state of a system whose two ends are kept at different
temperatures. Some of the important problems are to
see if local thermal equilibrium is achieved in the steady state, to
calculate the temperature profile inside the system and to 
check the validity of linear response laws relating heat current and
temperature gradient (Fourier Heat Law). In this context models with
both deterministic 
and stochastic dynamics have been studied. Rieder et al exactly solved
the problem of heat conduction in harmonic chains \cite{ried}. They
found a 
constant temperature distribution inside the system and a heat current
that was independent of system size. Since then, various authors have
studied the effect of impurity and nonlinearity 
in one and two dimensional systems of interacting atoms evolving via
deterministic Newtonian dynamics \cite{rubi,conn,casa}. Numerical
studies show that with some 
kinds of interaction one gets realistic temperature profiles and heat
currents proportional to inverse of system size \cite{casa}. For the
Lorenz gas in 
the Boltzmann-Grad limit, it has been proved that the linear
Boltzmann equation holds and in this case too one gets the Fourier Law
\cite{lebo}.
There are however not too many studies on heat conduction in
stochastic models.   
Connor and Lebowitz studied the random reflection model in which
particles moving on a line are reflected with some probability from
randomly spaced scatterers \cite{conn}. Streater has studied heat
conduction in a lattice model with finite degrees of freedom
\cite{stre}.  

In this paper we are interested in 
stochastic lattice models of heat conduction. Specifically we
look at classical spin systems with some kind of imposed dynamics. In
studying spin systems one typically uses dynamics such as 
Glauber where  each point of the lattice is kept in contact with heat
baths at given temperature. In this case the spin flip rates are
assumed to satisfy detailed balance condition, with the rates
depending on the heat bath temperature.  
For equilibrium studies the same temperature is imposed at all points
and the steady state is the canonical distribution. However, in
studying heat 
conduction, the temperature distribution in the steady state is not
known beforehand and should come out of the dynamics. Since the
temperature is space dependent and not known in the beginning, one
cannot specify the relaxation rates for energy raising and energy
lowering transitions. The only simple choice is to make such rates
zero.
Thus we use dynamics where there is no exchange of heat with
external reservoirs (except at the two ends) and there is local
conservation of energy. However, in this case the dynamics becomes
highly constrained, and it is not clear that it wont get stuck. A
requirement of correct phenomenological 
description is that the system should relax to a state of local
thermal equilibrium. In order that this occurs, we expect 
that the isolated system should reach thermal equilibrium, that is the
dynamics should be ergodic and lead to a microcanonical distribution. 
In this paper we address this problem only. We study heat conduction
in our model in a future paper. 

It is easy to show that the Ising model with single spin flip energy
conserving dynamics is non-ergodic in dimensions higher than one. For
example, suppose we take a 
$d$-dimensional hypercubic lattice of $N$ sites, and consider the spin
configuration of one single up-spin in a sea of down-spins. If we take
periodic boundary conditions then there are $N$ degenerate energy
configurations corresponding to the $N$ places where the single
up-spin can be located. The dynamics does not connect these states.
For other energies too one can show that there is no ergodicity.
In order to do microcanonical simulations for the Ising model it is
thus necessary to couple it to other degrees of freedom with which it
can exchange energy. This has been studied by Bhanot et al
\cite{bhan}. Heat conduction in the Ising model (with additional
degrees of freedom) was studied by Creutz \cite{creutz}.

One can ask what happens if we replace the Ising spins by continuous
spins. Microcanonical simulations of the $XY$ model have earlier been
done by introducing additional degrees of freedom. This include the
molecular dynamics study by Kogut and Polonyi who introduce momentum
variables and use a continuous time Hamiltonian dyanmics
\cite{kogut}. A microcanonical simulation using the Creutz algorithm
was done by Ota et al \cite{ota}.

In this paper we consider an energy conserving single spin flip
dynamics for the $XY$ model, with no additional degrees of freedom, and
look at its ergodicity properties.  
We first study the instructive case of two spins interacting with each
other and with external fields. This case is still non-trivial but can
be worked out in full detail. We show that for
generic values of external fields, the phase space trajectories (note
that the trajectory is not continuous and is formed by the random
iterates of a phase point) densely
fill the geometrically connected parts of the constant energy
contours. Thus whenever the energy contours are fully connected we
have ergodicity. On the other hand, even though the dynamics is
discontinuous, we show that jumps between
closed disconnected parts of the energy contours are not allowed.  
Thus in such cases we have non-ergodic behavior and we have as many
sectors as disconnected energy parts, within each of which the
dynamics is ergodic. Thus ergodicity is related to nature of the constant
energy contours.
Information regarding the nature of the
energy contours can be obtained from a study of the critical
points. These are points at which the derivative of the energy w.r.t
all angles vanish. For instance, if there are two or more stable
critical points (these correspond to fixed points of the dynamics)
with the same 
energy values then we will have closed energy contours around each of
them, which are disconnected. We find that the
nature of energy contours depend on the external fields. For some
fields there are only single sectors at all energies while for other
field values there are one or two sectors depending on the energy.

Next we study the case of $N > 2$ spins. 
Using the two spin results, we prove that even for the general
case of $N$ interacting spins, the phase space 
trajectory densely covers the geometrically connected parts of the
constant energy surface. Thus ergodicity is related to the
connectivity properties of the energy surfaces. 
As in the two-spin case, some information can be obtained by looking
at the critical points. As an illustrative example, we 
study a system of three spins interacting with each other and with
external fields.
We show that the energy surfaces can be constructed by
studying the critical points. In this case all energy surfaces are
connected and phase space trajectories densely fill the surfaces.
Finally we study spins on hypercubic lattices in $d$
dimensions. In this case we are not able to obtain a full
characterization of the energy surfaces. However we explicitly
construct a large number of spin configurations which correspond to
critical points, both stable and hyperbolic, and which span the entire
energy spectrum of the system. The presence of degenerate stable
points implies disconnected energy surfaces around the neighborhood of
these points. Thus in such cases we get non-ergodic behavior. 
However we find some numerical evidence which
suggests that such regions where the dynamics is non-ergodic are of
small measure. For example, we find from simulations using the energy
conserving dynamics that long time
averages of correlation functions are the same as that obtained from a
canonical simulation (which can be shown to be ergodic) 
whenever the initial conditions are chosen randomly. 
For initial conditions close to stable points the correlation
functions can be calculated exactly and differ from the canonical
averages. On the other hand if we start with initial
conditions close to hyperbolic points we get from simulations results
close to the canonical ones.

The paper is organized as follows. In section II we define the
model and the dynamics. 
In section III the two spin case is first studied. 
The general case of $N > 2$ spin case is studied in section IV. 

\vspace{1cm}

\section{Definition of model}

In the $XY$ model each spin is a two-component unit vector
specified by the angle $\theta$ that it makes
with some fixed axis. Equivalently every spin can be denoted by the
complex number $z=e^{i \theta}$. The interaction energy for spins on a
lattice is given by 
\begin{eqnarray}
{\cal{H}}=-\sum_{<i,j>} \cos{(\theta_i-\theta_j)}
\end{eqnarray}
where the sum is over all nearest neighbors.
The dynamics is the following: 
during an infinitesimal time $dt$ each spin undergoes an energy
conserving flip with a probability $dt$. For discrete time evolution, 
as would be required for a computer implementation of the dynamics,
this is equivalent to choosing a spin randomly at every time step and
flipping it. 
Let $h=\sum_{i \in n.n} z_i$ be the net 
magnetic field at a site due to all the other nearest neighbor sites.
Then the spin flip is given by the map
\begin{eqnarray}
z'=hz^*/h^*.
\end{eqnarray}
which corresponds to a reflection of the spin $z$ about the
direction $h$.

For ergodic behavior it is necessary that starting from a generic
initial condition it should be possible to access all other points on
the constant energy surface. The detailed balance condition then
ensures that a uniform probability distribution over the energy
surface is the stationary distribution. 

We note that a generalization of the above dynamics to the $n$-vector
model can be made. For the Ising model $(n=1)$ a flip can only occur
if a spin has equal number of up and down neighbors.
For the $XY$ model $(n=2)$ there is always a unique flipped
state. For higher $n$ the spin can flip to a continuum of values.     

\section{Ergodicity in two spin systems}

Consider two spins $z_1$ and $z_2$ coupled to each other by
the xy-interaction and also interacting with constant external fields
of magnitude 
$h_1$ and $h_2$ and directions given by the angles $\phi_1$ and
$\phi_2$. The fields $h_1$ and $h_2$ could be the effect of other
spins held fixed in a bigger system. The energy of the system is given by  
$$ {\cal{H}} = -h_1 cos(\theta_1-\phi_1)-h_2
cos(\theta_2-\phi_2)-cos(\theta_1-\theta_2).$$
Here $\theta_1$ and $ \theta_2$ are the angles of the two spins.
We can choose the reference axis such that $\phi_1=0$ and $\phi_2=\phi$.  
If we are interested only in finding the region that 
is accessed by the system , then for the two spin case,
we may as well flip the spins successively. This is
sufficient since flipping a spin twice successively leaves it unchanged.
 Thus we study the deterministic maps 
\begin{eqnarray}
z_1'=\frac{h_1+z_2}{h_1 + 1/z_2} {1}/{z_1}, ~~~~~~~
z_2'=\frac{h_2 e^{i \phi}+z_1'}{h_2 e^{-i \phi} +1/z_1'} {1}/{z2}
\end{eqnarray}

We first note the following points. In general any constant $\theta_i$
line will intersect the constant energy contour given by
${\cal{H}}=\epsilon$ at two points only. These correspond to the two
equal energy configurations of a spin in a net magnetic field. The
energy contours can thus be of the forms shown in fig. 1. In (a) 
the contours are connected while
in (b) and (c) they are disconnected. However it is clear that the
dynamics allows transitions between the disconnected parts in (b) but
not so in case (c). This is because, at every step of the dynamics, only
one spin can change. But moving between the disconnected parts of the
energy contours in (c) requires two spins to change simultaneously.
Thus whenever there are closed disconnected contours, that are
non-spanning, there can be no transitions from one to the other.

We now show that for generic fields and initial conditions the
trajectories are non-periodic and densely fill the constant energy contours. 
The proof consists of the following observations:

(i) Considered as a map in the $\theta_1$, $\theta_2$ plane, 
(3) is area preserving, i.e it has jacobian of modulus one. This is
easily proved by writing eqns. (3) in the form
$\theta_1'=-\theta_1+f(\theta_2), \theta_2'=-\theta_2+g(\theta_1')$,
where $f$ and $g$ are given functions. 
Thus the map can have no attractive periodic points.

(ii)The map (3) is equivalent to a smooth invertible map on the
circle. Since there can be no attracting set of points, two kinds of
behavior of the trajectories is possible. Either all trajectories are
periodic with the same period or else trajectories are aperiodic and
densely fill the circle.

(iii)For periodic orbits of order $n$ we must have 
$z_1^{(n)} = z_1$ and $z_2^{(n)} = z_2$. For generic values of the
fields, these equations will have a finite number of solutions, and
therefore it follows that all points cannot belong to periodic orbits.
Hence, because of (ii), all trajectories are aperiodic and fill the energy
contours densely. It is however possible that at some special value of
the fields the above eqns for periodic orbits are satisfied
identically for some $n$. In that case all trajectories are periodic
orbits of order $n$. Thus, as an example, for $h_1=h_2=1$, the
equations $z_1^{(3)}=z_1, z_2^{(3)}=z_2$ are identically satisfied. All
points are then periodic of order $3$ and there are is no
ergodicity. However if we change $h_2$ by an infinitesimal amount 
then the trajectories become dense. This can be seen in fig. 2
where we plot phase space trajectories for different initial
conditions for $h_1=1.0,h_2=1.001$.

For the two spin case the nature of the constant energy contours can
be completely determined by studying the critical points of the 
energy function
${\cal{H}}=-h_1 \cos(\theta_1)-h_2 \cos(\theta_2-\phi)-\cos(\theta_1-\theta_2)$.
The critical points are obtained by setting the first derivatives of
${\cal{H}}$ w.r.t $\theta_1$ and $\theta_2$ to zero. Thus we get:
\begin{eqnarray}
\frac{\partial
{\cal H}}{\partial \theta_1} &=& h_1 \sin(\theta_1)+\sin(\theta_1-\theta_2)=0
\nonumber \\ 
\frac{\partial
{\cal H}}{\partial \theta_2} &=& h_2
\sin(\theta_2-\phi)-\sin(\theta_1-\theta_2)=0 \nonumber \\   
\end{eqnarray}
These equations can be written in the more transparent form:
\begin{eqnarray}
z_1^2=\frac{h_1+z_2}{h_1+1/z_2} \nonumber \\
z_2^2=\frac{h_2 e^{i \phi}+ z_1}{h_2 e^{-i \phi}+ 1/z_1}. 
\end{eqnarray}
This form also follows from eqns (3) and imply that at the critical
points either the field at any site vanishes or is parallel to the spin
at that site. If the field does not vanish at both the sites then the
critical point will be a fixed point of the dynamics.
The nature of the critical points can be found by finding the
eigenvalues of the matrix $ M_{kl}={\partial^2
{\cal{H}}}/{\partial \theta_k \partial \theta_l}$ evaluated at the
critical points. 
We are interested in finding the phase diagram in the
($h_1,h_2,\phi)$) space showing regions where the energy contours
are connected and the regions where they are disconnected. As we shall
show now this can be obtained by studying solutions of eqn. (5).
First we note certain symmetries of eqn. (5). It is left
unchanged by the following transformations:
(1)$h_1 \to -h_1$, $z_2 \to -z_2$; (2) $h_2 \to -h_2$, $z_1 \to -z_1$;
(3)$z_1 \to 1/z_1$, $z_2 \to 1/z_2$, $\phi \to -\phi$; (4) $\phi \to
\pi-\phi$, $z_1 \to -1/z_1$, $z_2 \to 1/z_2$; (5) $h_1 \to h_2$, $h_2
\to h_1$, $ z_1 \to e^{i \phi}/z_2$, $z_2 \to e^{i \phi}/z_1$ . It can
also be shown that the equations determining the nature of the critical
points are also invariant under the above transformations.  Because of
these symmetries it is sufficient to consider the subspace given by
$h_1 > h_2 > 0, 0 < \phi < \pi/2 $.

From eqn (5) we get the following equation for $z_1$
\begin{eqnarray}
h_1^2 h_2 e^{-i \phi} z_1^6 + (h_1^2 h_2^2+h_1^2-h_2^2 e^{-i 2 \phi})
z_1^5 + (h_1^2 h_2 e^{i \phi}-2 h_1^2 h_2 e^{-i \phi}) z_1^4 +( -2  
h_1^2 h_2^2-2 h_1^2+2 h_2^2) z_1^3 + \nonumber \\ 
(h_1^2 h_2 e^{-i \phi}-2 h_1^2 h_2 
e^{i \phi}) z_1^4 + (h_1^2 h_2^2+h_1^2-h_2^2 e^{i 2 \phi}) z_1 + h_1^2 
h_2 e^{i \phi}=0
\end{eqnarray}
For every solution $z_1$ of (6), $z_2$ is uniquely determined through the
equation:
\begin{eqnarray}
z_2=\frac{h_2 z_1 ( e^{i \phi}-e^{-i \phi} z_1^2 )}{h_1 (z_1^2-1) (h_2
e^{-i \phi} z_1 +1)} \nonumber
\end{eqnarray}
Thus in general we obtain six sets of solutions to (5). The solutions
for which $z_1$ and $z_2$ have modulus one correspond to the physically
relevant solutions.

Let us first consider the case $\phi=0$, for which (6) can be solved
for $z_1$. We obtain the following six solutions:
$(z_1,z_2)=(1,1),(1,-1),(-1,1),(-1,-1),(e^{i \theta_1},e^{-i
\theta_2})$ and $ (e^{-i \theta_1},e^{i\theta_2})$ where
$\cos(\theta_1)= (h_2^2-h_1^2-h_1^2 h_2^2)/{2 h_1^2 h_2}$ and
$\cos(\theta_2)= (h_1^2-h_2^2-h_1^2 h_2^2)/{2 h_2^2 h_1}$.
We see that two of the critical points merge and disappear as the lines
$h_2=h_1/(h_1 \pm 1), h_2=h_1/(1-h_1)$ are
crossed. In fig. 3 we plot the phase diagram which shows the nature of
the critical points in different regions in the $h_1, h_2$ plane. The
important point to note is that in the region enclosed by the three
curves, where we have
three stable fixed points, we get disconnected energy contours. This
is because of the following reason. Consider what happens as we move
along some constant $h_1$ line as shown in fig. 3. At point $A$ just
below the phase boundary we have two stable critical points;$(1,1)$,
which is a energy minima and $(-1,1)$ which is a maxima. The other
two points $(-1,-1)$ and $(1,-1)$ are hyperbolic points. As we cross
the line to point $B$, $(-1,1)$ becomes unstable and at the same time,
two new stable points, degenerate in energy appear from this
point. The separatrix passing through $(-1,1)$ encloses both the
stable points. This at
once implies that there will be disconnected closed energy contours of
identical energies around the two stable points. In fact for all
energy values greater than the separatrix energy, we get disconnected
sectors, while for lower energies there are only single
sectors. Similarly as we cross from the point $D$ to $E$ in fig. 3, the
two degenerate stable maxima merge with the unstable point $(-1,-1)$
to give rise to a single stable point. Hence at $E$ and all points
above this we have only connected energy contours for all energy
values. To illustrate the above
point we plot in fig. 4 the energy contours for field values
corresponding to the points $A,B,C,D$ and $E$. 
The energy contours were obtained by starting from different
initial conditions and evolving through the map (3).

For $\phi \ne 0$, we can no longer solve eqn (6) explicitly. However
we can still 
obtain the phase diagram. In this case for fixed $h_1$ suppose we
increase the $h_2$ value starting from zero. Initially we have 
two stable critical points, one of which is a minima and the other a
maxima, and two hyperbolic points. As we increase $h_2$ 
a pair of critical points, one stable and the other
hyperbolic, appears in the region around the maxima. The separatrix
passing through this new hyperbolic point encloses both the maxima and
the new stable point which is also a maxima. The two maxima may not be
degenerate but clearly, for a range of energy values greater than the
separatrix energy and less than the energy of the smaller maxima, we
will have disconnected energy contours. Since the extra pair of critical
points emerge from a single point,  
the phase diagram can be obtained by finding the
condition for two roots of eqn (6) to be identical. The equation that
we thus get can be solved numerically and we obtain the phase diagram
shown in fig. 5. This shows various constant $\phi$ sections of the
phase diagram. In fig. 6 we plot constant energy contours for
non zero $\phi$ and for field values on the two sides of the phase
boundary. The plot illustrates the appearance of critical points and
associated with it the disconnected energy contours.

To summarize the results obtained so far, we have shown that for a two
spin system, for generic values of external fields, the dynamics is
ergodic on the geometrically connected parts of energy contours. There
are as many sectors as disconnected parts. Studying the critical
points enables us to obtain full information on the number of sectors
as a function of the fields and energy. 

\section{ Ergodicity in an N spin system}

We now give some results that are valid for any number of spins. 
In this case too, if there are closed disconnected surfaces with the
same energy, the dynamics does not allow transitions between them.
This simply follows from the fact that any line, along which only
$\theta_k$ varies, 
will intersect a closed energy surface at two points. Since we allow
single flips only, at every time step, one can move only between these
two points on the same surface.

It can also be shown that for the many spins case, 
phase space trajectories obtained for generic initial 
conditions densely fill the connected parts of the constant energy
surface. This is an important result since it implies ergodicity
within sectors. 
Consider any point $P(\theta_1,\theta_2,...\theta_N)$ on the
energy surface. Suppose we take any two nearest neighbor spins
$\theta_k$ and $\theta_l$ and flip
them successively while keeping all other spins fixed. The fixed spins in
this case act as external fields. We know from the two
spin case that for almost every value of the
external fields, which in this case means for almost any configuration
of the fixed spins, the trajectory will be dense in the restricted
phase space. The restricted phase space is the intersection between
the energy surface and the $\theta_k$-$\theta_l$ plane. By choosing
different nearest neighbor pairs of spins, we can have 
$(N-1)$ independent directions emerging from the point $P$ on the
$(N-1)$ dimensional energy surface, along each of which there exists a
dense orbit. This in turn means that, for every point there exists a
neighborhood such that all points of the neighborhood can be reached
by the dynamics. Hence the trajectory obtained by evolving any initial
condition cannot have a boundary and so has to fill all of the
connected part of the energy surface.

Finally to obtain information on the connectivity of the energy
surfaces, one can study the critical points. However unlike the two
spin case, for more number of spins, we are not able to do this in
full detail. 

In the following subsections, first as an illustrative example, we treat
the case of three spins in external fields. Then we consider spins on
general hypercubic lattices.

\subsection{Ergodicity in Three-spin case}

For an open chain of three spins interacting with each other and with
external fields, the energy function is given by 
\begin{eqnarray}
{\cal{H}}=-h_1 cos(\theta_1-\phi_1)-h_2 cos(\theta_2)-h_3
cos(\theta_3-\phi_3)- cos(\theta_1-\theta_2)-cos(\theta_2-\theta_3)
\nonumber 
\end{eqnarray}

In this case depending on the the sequence in which the spins are
flipped we get different maps in the three dimensional phase space.
Here we look at the stochastic map in which any one spin is chosen
with equal probability and flipped. Thus if we represent the three
spins by complex numbers $z_1$, $z_2$ and $z_3$ then at every time
step we choose one of the following maps:
\begin{eqnarray}
z_1'=\frac{h_1 e^{i \phi_1}+z_2}{h_1 e^{-i \phi_1}+1/z_2} {1}/{z_1}
\nonumber \\
z_2'=\frac{h_2+z_1+z_3}{h_2+1/z_1+1/z_3} {1}/{z_2} \nonumber \\
z_3'=\frac{h_3 e^{i \phi_3}+z_2}{h_3 e^{-i \phi_3}+1/z_2} {1}/{z_3}
\end{eqnarray}

Let us study this dynamics for the special case of fields given by:
$h_1=h_2=h_3=1.0$ and $\phi_1=\phi_3=0.0$. We first look at the
nature of the constant energy surfaces. For this we study the
critical points.

The critical points are obtained by
setting to zero the first derivatives of $\cal{H}$ w.r.t
$\theta_1$,$\theta_2$ and $\theta_3$. This is equivalent to
the condition that at every site the net field is either parallel to
the spin at that site or it vanishes. Thus from (7) we get:
\begin{eqnarray}
z_1^2=\frac{1+z_2}{1+1/z_2} \nonumber \\
z_2^2=\frac{1+z_1+z_3}{1+1/z_1+1/z_3} \nonumber \\
z_3^2=\frac{1+z_2}{1+1/z_2} .
\end{eqnarray}
Solving the above equations we obtain the following critical points:
$P_1(1,1,1)$, $P_2(-1,1,-1)$, $P_3(1,1,-1)$, $P_4(-1,1,1)$,
$P_5(z_1,-1,1/z_1)$, $P_6(z_1,-1,-z_1)$. 
In fig. 7, we show the locations of these critical points in the
$(\theta_1,\theta_2,\theta_3)$ phase space.

By expanding 
${\cal H}$ around the critical points it can be shown that $P_1$
is a minimum of the energy with $\epsilon=-5$ while $P_2$ is a maximum
with $\epsilon=3$. 
The points $P_3$ and $P_4$ are hyperbolic points with $\epsilon=-1$. 
The lines $P_5$ and $P_6$ consist of hyperbolic points and correspond to
$\epsilon=1$. 
Knowing the nature of all the critical points and by carefully looking
at the energy function, we 
construct the following picture of the energy surfaces:
The point $P_1$ is the lowest energy state corresponding to all spins
aligned along the field direction. For $\epsilon \approx -5$, the
constant energy surfaces are ellipsoidal, enclosing the point $P_1$.
In fact for  $-5 < \epsilon <-1$ the energy surfaces are all closed
deformed ellipses.
At $\epsilon = -1$ the energy surface passes through the two unstable
points $P_3$ and $P_4$. For $ -1 < \epsilon < 1$ we get spanning surfaces.
The $\epsilon =1$ surface is again  closed  and passes through the
critical lines $P_5$ and $P_6$. Finally for $ 1 < \epsilon < 3$, the
surfaces are closed around the other stable point at $P_2$.
We note that in all cases the energy surfaces are connected.

Let us now look at the numerically obtained phase space trajectories
for different energy values. These are shown in fig. 8. For every
energy value the trajectory was obtained by evolving a single initial
condition. In all cases the dynamics appears to be ergodic. The energy
values were chosen so that all the different types of surfaces
discussed above are generated. For example it can be seen that at
$\epsilon=-0.832$ we get a spanning surface while for $\epsilon > 1$ the
surfaces are closed. The surface in fig. 8c is a spanning surface
topologically equivalent to the one in fig. 8d. To further verify the
ergodic nature of the 
trajectories we plot in fig. 9 constant $\theta_2$ sections of the
phase space trajectory for the 
energy value $\epsilon=-0.832$. It is easy to understand the form of
the contours if we rewrite the energy in the following way
\begin{eqnarray}
{\cal H}=2 \cos(\theta_2/2) ( \cos(\theta_1-\theta_2/2 )+
\cos(\theta_3-\theta_2/2) ) +\cos(\theta_2) \nonumber .
\end{eqnarray}
For other values of energy we get results similar to those in fig. 9.

Thus in this three spin system we find that all energy surfaces are
connected and the phase space trajectories densely fill the energy
surfaces. Hence we have complete ergodicity.

\subsection{Spins on a lattice}

On a general hypercubic lattice consisting of $N$ sites 
the equations for the critical points is 
\begin{eqnarray}
z_i=\frac{\sum_{j \in <i>} z_j}{\sum_{j \in <i>} 1/z_j}
{1}/{z_i}~~~~~~~i=1,2.....N. 
\end{eqnarray}
The sum over $j$ is over nearest neighbors of $i$.  In all cases we
assume periodic boundary conditions. We have not been
able to obtain the full set of solutions of these equations and
analyze their structure. We have done this for a class of solutions
and this is described below. The one dimensional case is a special
case and we first describe it. 

\vspace{0.5cm}
{ \bf{(i) One dimensional lattice}}
\vspace{0.25cm}

First we note that in this case, if we set all the coupling constants
strictly to one then the dynamics is not ergodic even within sectors.
The proof of ergodicity in III breaks down.
This is because fixing all spins except two nearest neighbors now
corresponds to the special case $h_1=h_2=1.0$ mentioned in section III,
which was shown to be periodic.
However it was seen that if we make the couplings different from one
by arbitrarily small amounts, then 
we get ergodicity (within sectors). Assuming this has been done
we can still determine the phase space structure with the couplings set
equal to one.

The set of eqns. (9) can be written in the following equivalent form:
\begin{eqnarray}
z_k^2 = z_{k-1}z_{k+1}~~~~\rm{or}~~~~ z_{k-1}+z_{k+1}=0~~~k=1,2...N.
\end{eqnarray}
Due to periodic boundary condition we have $z_0=z_N$ and $z_{N+1}=z_1$.
Thus we have $2^N$ sets of equations whose solution would give us the 
critical points. It is easy to see that the only stable solutions
(stable in all directions except for a uniform rotation) are obtained
from the set of equations 
\begin{eqnarray}
z_k^2=z_{k-1}z_{k+1}~~~~~ k=1,2..N. 
\end{eqnarray}
Every
other of the $2^N$ sets gives saddle points since there are one or
more sites where the field vanishes and at those sites one can freely
rotate the spins without changing the energy. For this set since the
fields don't vanish at any site they also correspond to fixed points of
the dynamics. That the above set is
stable is proved by explicitly solving for the fixed points
and analyzing their nature. We get $N$ solutions given by $z_k(n)=e^{2
\pi i k n/N} z_1$ with $n$ going from $1$ to $N$. The factor $z_1$
signifies the fact that the direction of one of the spins can be fixed
arbitrarily. We now look at the matrix $M_{kl}={\partial^2
{\cal H}}/{\partial \theta_k \partial \theta_l}$ evaluated at the fixed
points. The non-vanishing matrix elements thus obtained are
$M_{kk}(n)=-2\cos(2\pi 
n/N)$, $M_{k k-1}=M_{k k+1}=cos(2 \pi n/N)$. 
Here $n$ refers to the $n$th fixed point. This can be easily
diagonalised and gives the eigenvalues 
$$\lambda_q(n)=-2 \cos(2 \pi n/N) (1-\cos(2 \pi q/N))~~~~~q=0,1,...N-1.$$
Except the first eigenvalue $q=0$ which corresponds to a uniform
rotation, all other eigenvalues are non-vanishing and have
the same signs. Thus all fixed points are stable and the energy
surfaces around them are elliptical. For the case when
$N$ is a multiple of $4$, a special case arises for the $n=(N/4)$th fixed
point. In this case all eigenvalues vanish. This is because in this
case the field at every site vanishes and spins can rotate freely.

The energy of the $n$th fixed point is given by $\epsilon=N \cos(2 \pi
n/N)$ and we note that every fixed point except those corresponding to
the lowest and highest energies are doubly degenerate (for even $N$).
The energy surfaces around these fixed points thus consist of
two disconnected parts each of which is closed.

Another set of solutions of eqns.(10) is obtained by choosing any
arbitrary direction and aligning every spin either along or opposite
to it. The all up  and the alternating up and down configurations 
have already been generated in the previous set of solutions. All
the other solutions are saddles and it is easy to see that the
degenerate points are all connected. The degeneracy of each solution is
given by the number of configurations with a fixed number of bad bonds. 

In summary, for the one dimensional lattice, we have obtained all the
fixed points and shown that they are stable and doubly degenerate.

\vspace{0.5cm}
{\bf{(ii) General case of $d$-dimensional hypercubic lattice of length $L$}}
\vspace{0.25cm}

In higher dimension, eqns. (9) are highly nonlinear and classifying
all solutions is a difficult task. However we obtain by inspection a 
set of solutions corresponding to spin-waves. These are of the form 
\begin{eqnarray}
z_{\bar k}(\bar n)=e^{2 \pi i{\bar k}.{\bar n}/L}{z}
\end{eqnarray}
where the vector $\bar k$ denotes points on the d-dimensional lattice
with integral coordinates $(k_1,k_2,...k_d)$ and $\bar
n=(n_1,n_2,...n_d)$ denotes one of the $L^d$ solutions. The indices
$k_i$ and $n_i$ take values $1,2,...L$. As before the solutions
contain an arbitrary constant phase factor $z$.

To understand the nature of the critical points we again look at the
eigenvalues of the matrix $M_{\bar k \bar l}={\partial^2
{\cal H}}/{\partial \theta_{\bar k} \partial \theta_{\bar l}}$,
evaluated at the critical points. The non-vanishing matrix elements
are in general given by $M_{\bar k \bar k}=-\sum_{\bar l \in <\bar k>} 
\cos(\theta_{\bar k}-\theta_{\bar l})$ and $M_{\bar k \bar
k_p}=\cos(\theta_{\bar k}-\theta_{\bar k_p})$. Here $\bar k_p$ denotes the
nearest neighbors along the $p$th axis, namely the points
$(k_1,k_2,....k_p \pm 1,...k_d)$. In our case the matrix elements at the
critical point $\bar n$ are $M_{\bar k \bar k}(\bar n)=-2 \sum_{p=1}^d
\cos(2 \pi n_p/L)$ and $M_{\bar k \bar k_p}(\bar n)=\cos(2 \pi
n_p/L)$. It is easy to show that the eigenvalues of this matrix are
\begin{eqnarray}
\lambda_{\bar q}(\bar n)=-\sum_{p=1}^d 2 \cos(2 \pi n_p/L) (1-\cos(2
\pi q_p/L))~~~~~q_p=0,...N-1.
\end{eqnarray}
Unlike in the one-dimensional case eigenvalues corresponding to a
critical point can have opposite signs. In fact the only stable
critical points are those for which $\cos(2 \pi n_p/L)$ has the same
sign for all $p$. To show this suppose there exists in the set
$(n_1,n_2... n_d)$ two values $n_a$ and $n_b$ for which $\cos(2
\pi n_a/L)$ and $\cos(2 \pi n_b/L)$ have the opposite signs. Then
the eigenvalues for $\bar q=(0,0,...q_a,0,..0)$ and $\bar
q=(0,0,...,q_b,0,...0)$ have opposite signs. To find whether the
critical points correspond to fixed points we look at the fields at
each site. This is given by $H_{\bar k}(\bar n)=\sum_{\bar k_p}
z_{\bar k_p}(\bar n)=2 e^{2 \pi i {\bar k. \bar n}} \sum_{p=1}^d
\cos(2 \pi n_p/L)$. Thus except for the small subset of points
satisfying $\sum_{p=1}^d \cos(2 \pi n_p/L)=0$, all other critical
points are fixed points. Unlike the one dimensional case, here we may
have other fixed points corresponding to other solutions of eqns.(9). 

The energy of the critical point $\bar n$ is 
$\epsilon=N \sum_{p=1}^d \cos(2 \pi n_p/L)$. We note that the spectrum
of energies spans the entire allowed energy range. If $d_g$ is the number of
non-zero coordinate values in the set $\bar n$, then there is a
degeneracy of $2^{d_g}$ associated with the corresponding energy value 
since we can flip the signs of $n_i$ without affecting the energies.

The fact that there exist degenerate stable fixed points implies that
there are disconnected energy surfaces corresponding to the
neighborhoods of these points. This proves that the dynamics
is non-ergodic for certain energies. It is however possible that we can
get ergodicity for energies corresponding to unstable points and their
neighborhood. The following numerical simulations address this question. 

We took a $20 \times 20$ square lattice and computed $1$st, $2$nd,
 $3$rd and $4$th nearest neighbor correlation functions
($c(1)$,$c(2)$,$c(3)$,$c(4)$) using monte-carlo simulation. 
We study the two cases when the initial data is chosen close to a
stable or close to an unstable point. In each case we compare the
results with those obtained from randomly chosen initial configuration. 
Close to a stable fixed point the correlation function 
$<\cos({\theta_{\bar k}-\theta_{\bar k+\bar R}})>$
 can be calculated exactly. At the fixed point
$\bar n$ this is simply $\cos(2 \pi {\bar R. \bar n}/L)$. The
correlation $c(r)$ evaluated in the simulations corresponds to $(cos(2
\pi r n_1)+cos(2 \pi r n_2))/2$. The following table gives the
numerical data and also the exact values at the stable points:  
\begin{figure}
\begin{tabular}{|c|c|c|c|c|}  \hline 
Initial Condition  &~~~$\epsilon /2=$ c(1)  &~~~ c(2) &~~~ c(3) &~~~ c(4) \\ \hline
$\bar n \approx (2,6)$/Unstable       &~~~ 0.24855 &~~~ 0.06646 &~~~ 0.01861 &~~~ 0.00524 \\ \hline  
Random             &~~~ 0.24855  &~~~ 0.06647  &~~~ 0.01898 &~~~ 0.00574 \\ \hline
$\bar n \approx (3,4)$/Stable     &~~~ 0.44640 &~~~ -0.55508 &~~~ -0.87305 &~~~ -0.24811 \\ \hline
Random             &~~~ 0.44640 &~~~ 0.23264 &~~~ 0.13246 &~~~ 0.07996 \\ \hline  
Exact              &~~~ 0.44840  &~~~ -0.55902  &~~~-0.88004  &~~~ -0.25
\\ \hline
\end{tabular}
\end{figure}
As expected for the stable case we get results totally different from
the random case. On the other hand at the unstable points we get the
same correlations as 
with random initial conditions. Thus as far as correlation functions
are concerned there seems to be no dependence on initial conditions
provided we don't start from near stable points. 

Next we show that the correlation functions obtained with randomly
chosen initial conditions are the same as those obtained from a
canonical simulation.
In fig. 10 we plot the correlations $c(2)$, $c(3)$ and $c(4)$ against
the average energy given by $c(1)$ for
data obtained from the energy conserving dynamics and from a canonical
simulation. In the canonical simulation each spin is coupled to a heat
bath at some temperature and clearly we have ergodicity. 
For the energy conserving dynamics we started with a high energy
configuration of spins pointing in random directions. Subsequently the
energy 
was decreased by choosing spins randomly and aligning them along the
local field direction.  For each configuration obtained in this way we
calculated the correlation functions.
The data from the two approaches seem to match closely.

The above numerical results suggest that:

(1) the set of stable points and the sectors associated with them
seem to have a small measure.

(2) the unstable points with the same energies are either all
connected by the dynamics or correlation functions in each sector are
the same.

A more complete study of the phase space structure is needed in order
to understand the above results.

Finally we note that as in the one dimensional case, 
another class of solutions of the critical point equations can be
written. These are the solutions for which every spin points either
along or opposite to a fixed direction. These critical points may be
either stable or unstable. For example consider the spin configuration
of a single up-spin in a sea of down-spins. This is an unstable fixed
point. The different fixed points, corresponding to the $N$ different
locations of the single up-spin, are all connected. To show this
suppose we start from a configuration which is arbitrarily close to
such a fixed point. Now we fix all spins except for the single up-spin
and any one of its down neighbors. The results in section III imply
that by flipping these two spins successively we can reach the
configuration in which the two spins have interchanged their
orientations. Repeating this for different nearest neighbor spin pairs,
we can reach all the $N$ fixed points. This is different from the
Ising model with similar dynamics where the system would remain stuck in
one of the fixed points. 
The above procedure can be used to prove hyperbolicity of other fixed
points also.

In conclusion, we have studied a stochastic energy conserving single
spin flip dynamics of the $XY$ model. We find that phase space
trajectories fill connected 
parts of the energy surface and there is loss of ergodicity whenever
the energy surface consists of closed disconnected parts. 
We also show that one can obtain information about the nature
of the energy surfaces by studying the critical points of the energy
function. This can be done in full detail for a two-spin system. For
spins on hypercubic lattices we have proved that the dynamics is not
ergodic for all energies and time averages of physical quantities
depend on choice of 
initial conditions. However, the total weight of non-typical initial
conditions appears to be small and the dynamics is effectively ergodic.

I would like to thank Deepak Dhar for valuable suggestions all through
the course of this work.

{\centerline{\bf Figure Captions}}

\begin{figure}
\caption{Schematic picture of  constant energy contours for two spin case.~~~~~~~~~~~~~~~~~~~~~~~~~~~~~~~~~~~~~~~~~~~~~~~~~~~~~~~~~~~~~~~~~~~~~~~~~~~~}
\end{figure}
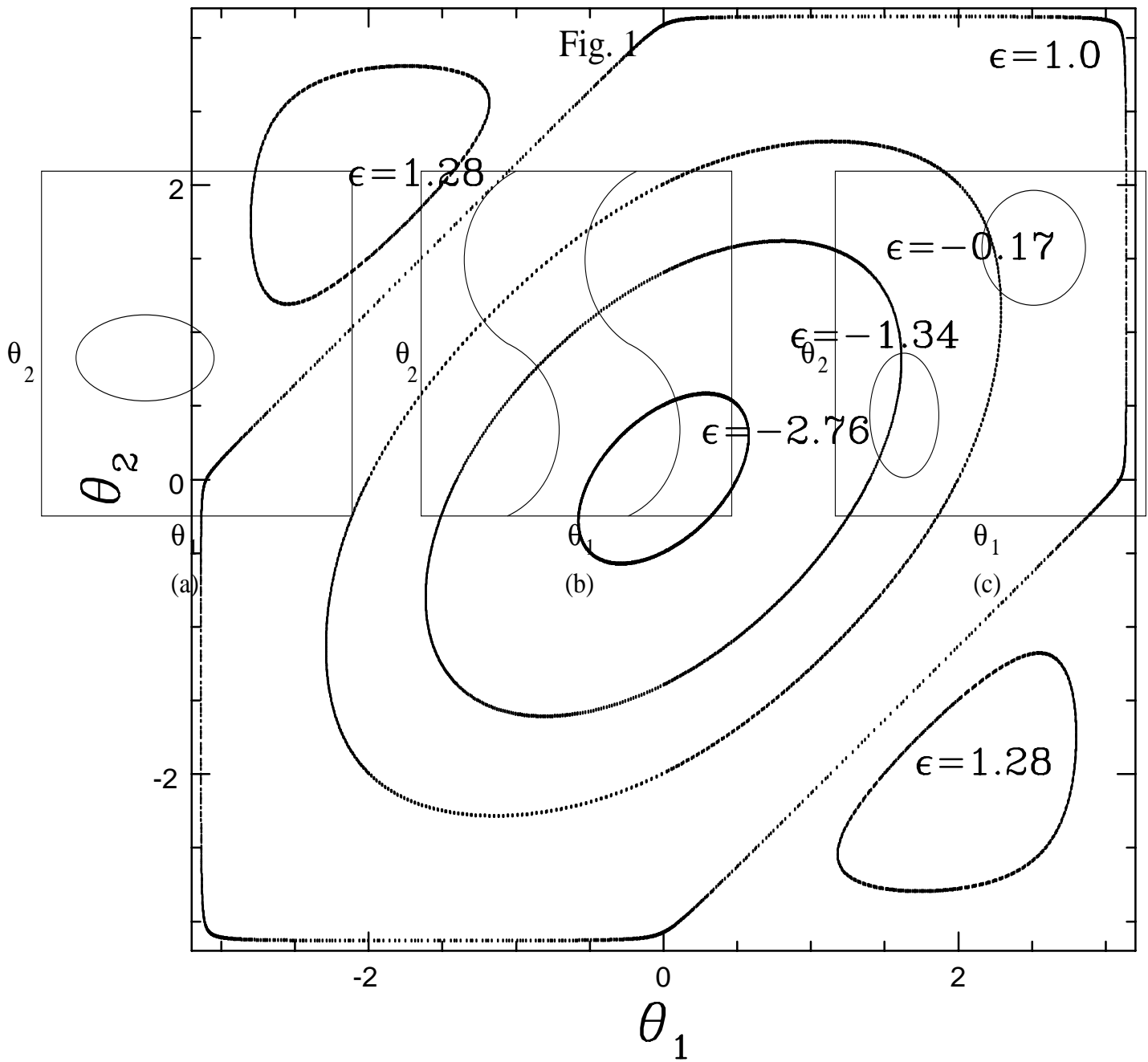
\begin{figure}
\caption{Phase space trajectories of the map (3) for $ h_1=1,
h_2=1.001$, and $\phi=0$.~~~~~~~~~~~~~~~~~~~~~~~~~~~~~~~~~~~~~~~~~~~~~~~~~~~~~~~~~~~~~~~~~~~~~~~}
\end{figure}
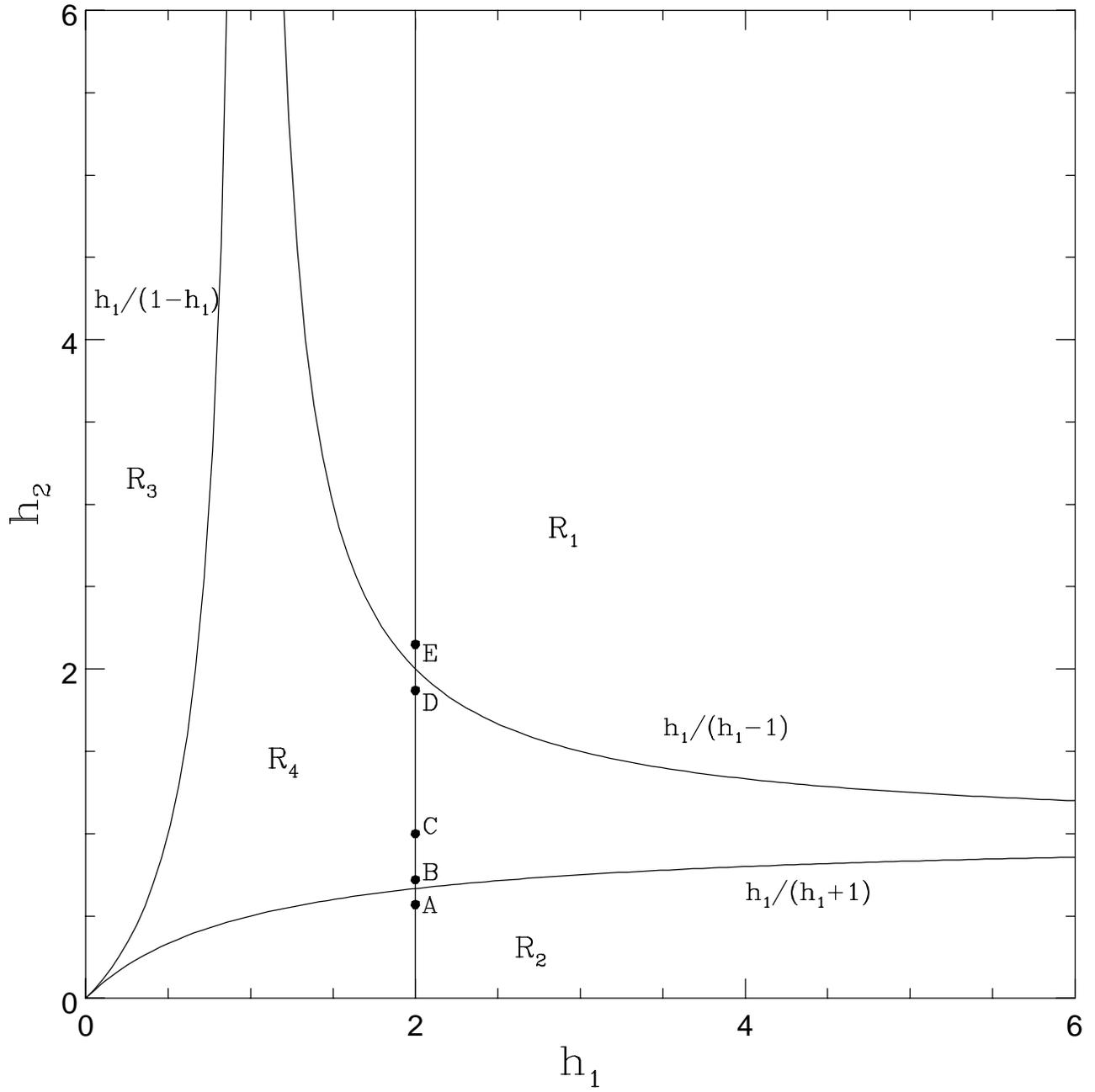
\begin{figure}
\caption{Phase diagram for $\phi=0$, showing nature of critical
points for different field values. In regions $R_1$, $R_2$ and $R_3$ there
are four critical points of which two are hyperbolic, one a maxima and
the other a minima. In region $R_4$ there are six critical points. Three
of them are hyperbolic, two are maxima and one is a minima.}
\end{figure}
\begin{figure}
\caption{Constant energy contours obtained by iteration of the map (3)
starting from different initial conditions. The figures (a)-(e) are
for five different field values corresponding to the points $A$-$E$ in
fig. 3. In all cases, some of the initial conditions were chosen close
to the hyperbolic points, so that the separatrices can be seen.} 
\end{figure}
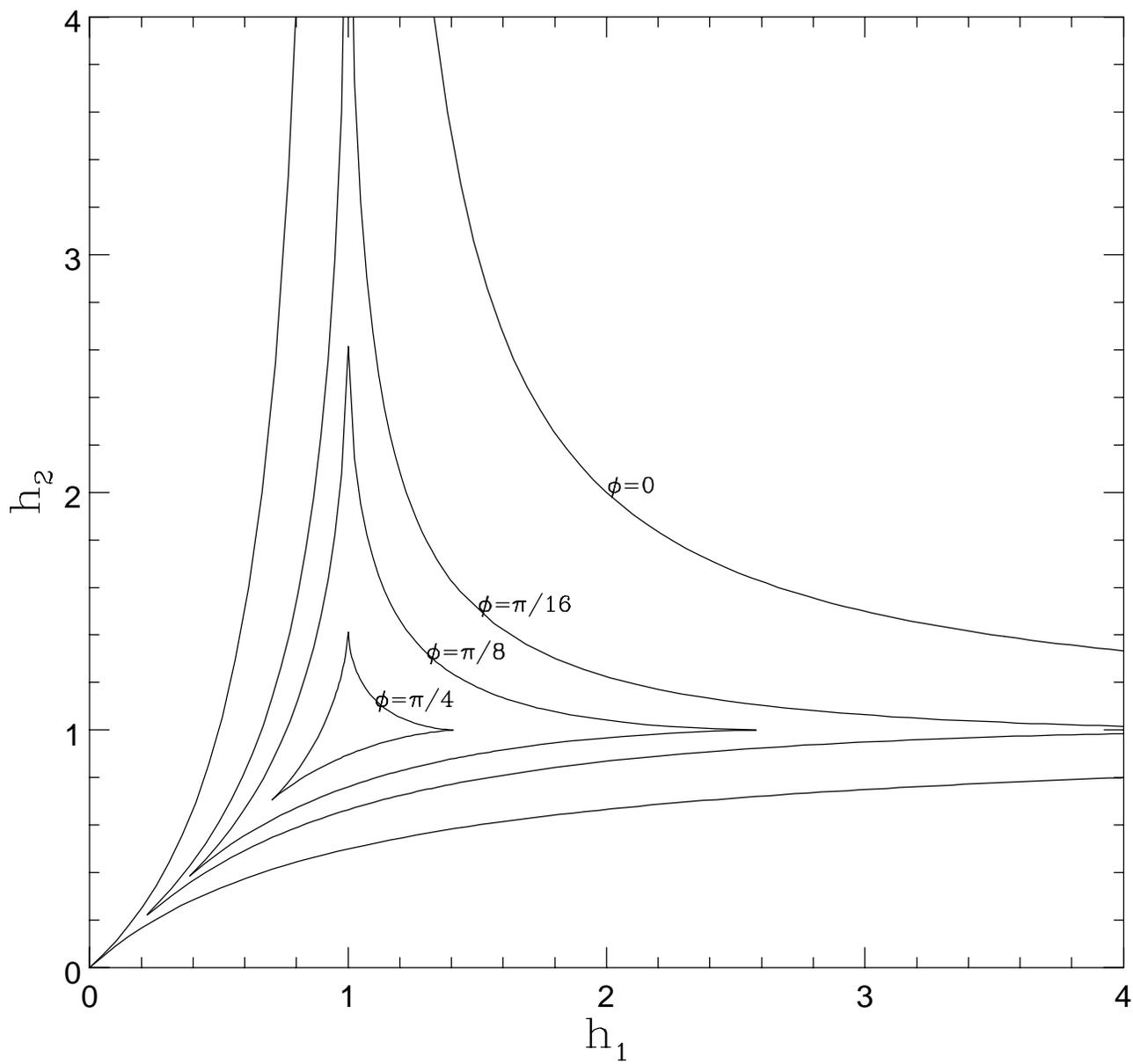
\begin{figure}
\caption{Phase diagram for different values of $\phi$. In
every case, for field values in the region bounded by the three curves,
there is a range of energy where there are two sectors.}
\end{figure}
\begin{figure}
\caption{Constant energy contours for $\phi=\pi/16$. The field
strengths $h_1$ and $h_2$ were chosen to lie on two sides of the phase
boundary shown in fig. 5. In (a) $h_1=1.25, h_2=0.77$ while in (b)
$h_1=1.25, h_2=0.89$. }
\end{figure}
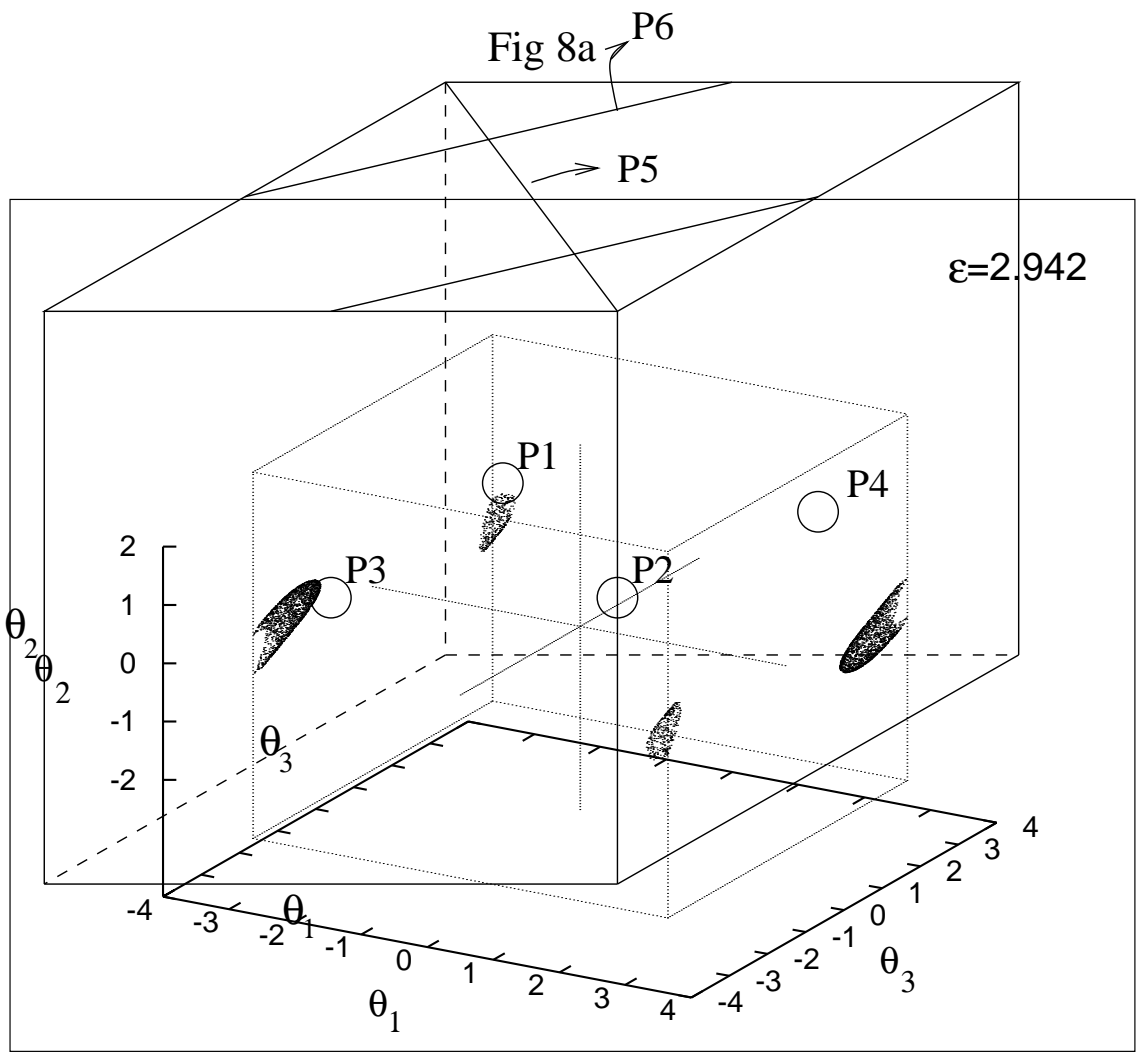
\begin{figure}
\caption{Position of critical points of three spin energy function in
the $(\theta_1,\theta_2,\theta_3)$ phase space. The value of $\theta$
ranges from $-\pi$ to $\pi$ along each of the
axes.~~~~~~~~~~~~~~~~~~~~~~~~~~~~~}  
\end{figure}
\begin{figure}
\caption{Three dimensional phase space trajectories for field values
$h_1=h_2=h_3=1.0$. Each of the $5$ figures was obtained by iterating
different initial conditions corresponding to different energy
values.} 
\end{figure}
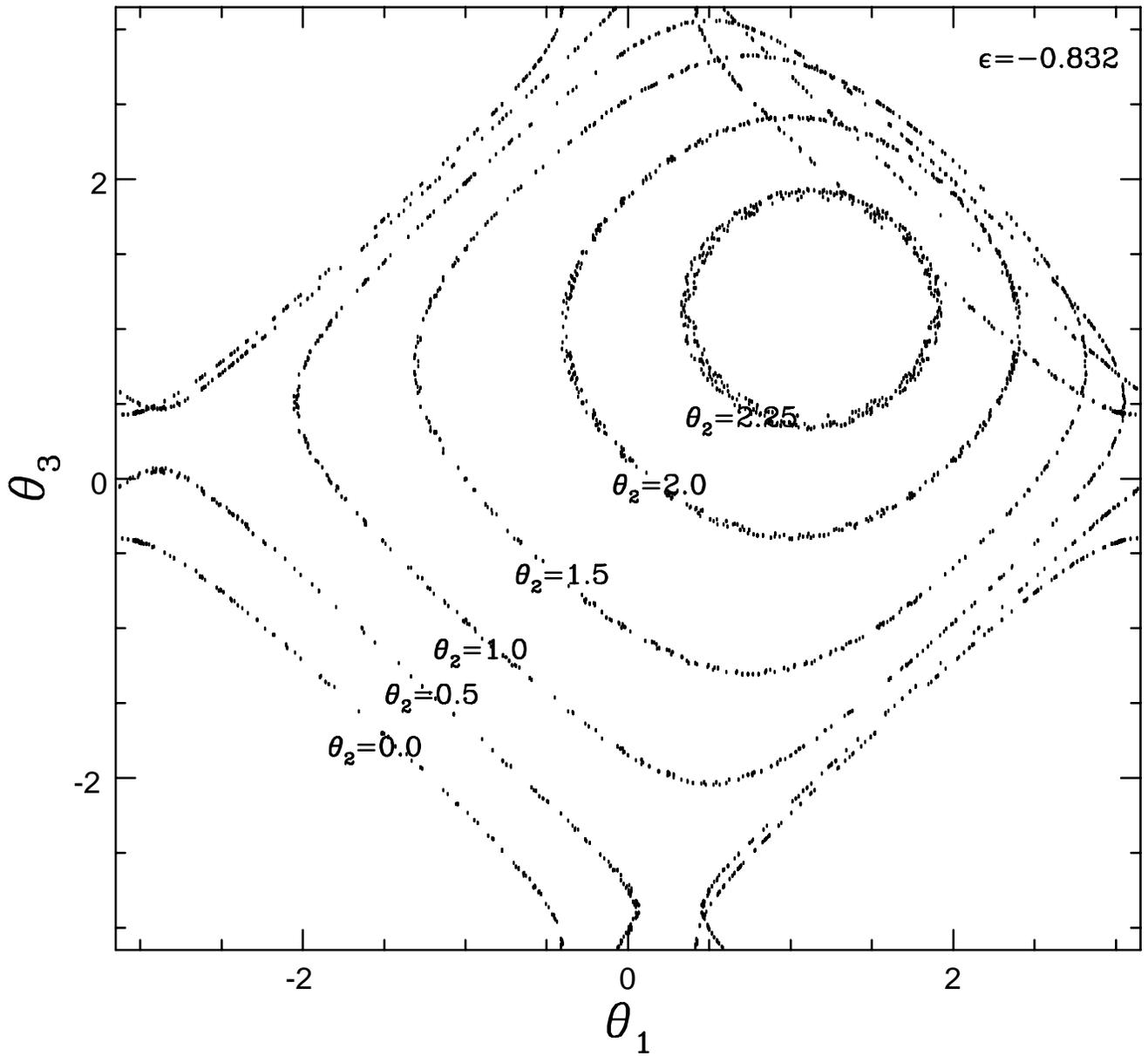
\begin{figure}
\caption{Constant $\theta_2$ sections of the phase space trajectories
for the energy value $\epsilon=-0.832$.~~~~~~~~~~~~~~~~~~~~~~~~~~~~~~~~~~~~~~~~~~~~~~~~~~~~}
\end{figure}
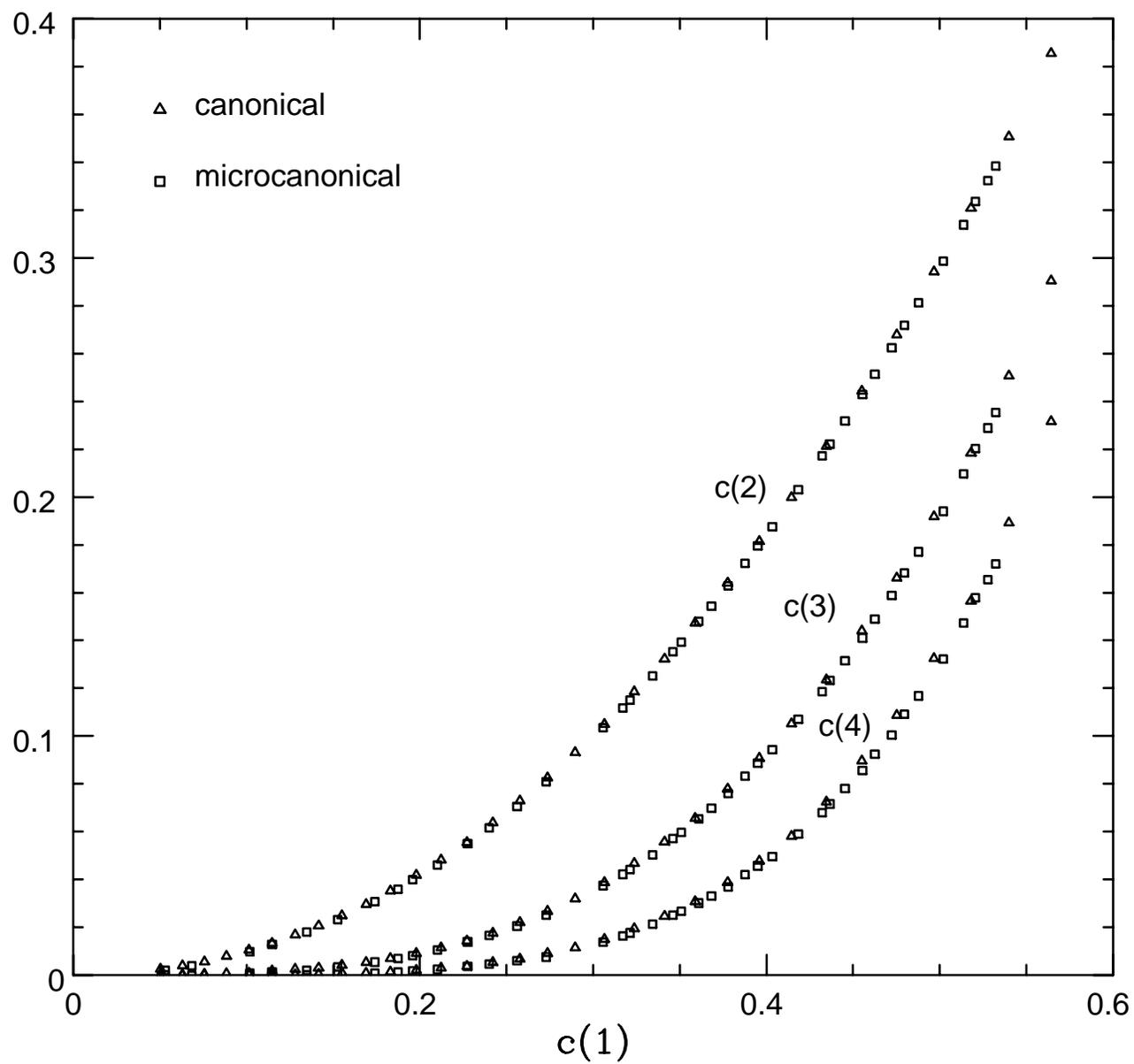
\begin{figure}
\caption{Correlation functions $c(2),c(3)$ and $c(4)$ plotted against
$c(1)$ for data obtained using energy conserving and non-conserving
dynamics. In each case the averages were taken over $10^5$
monte-carlo steps.} 
\end{figure}
 
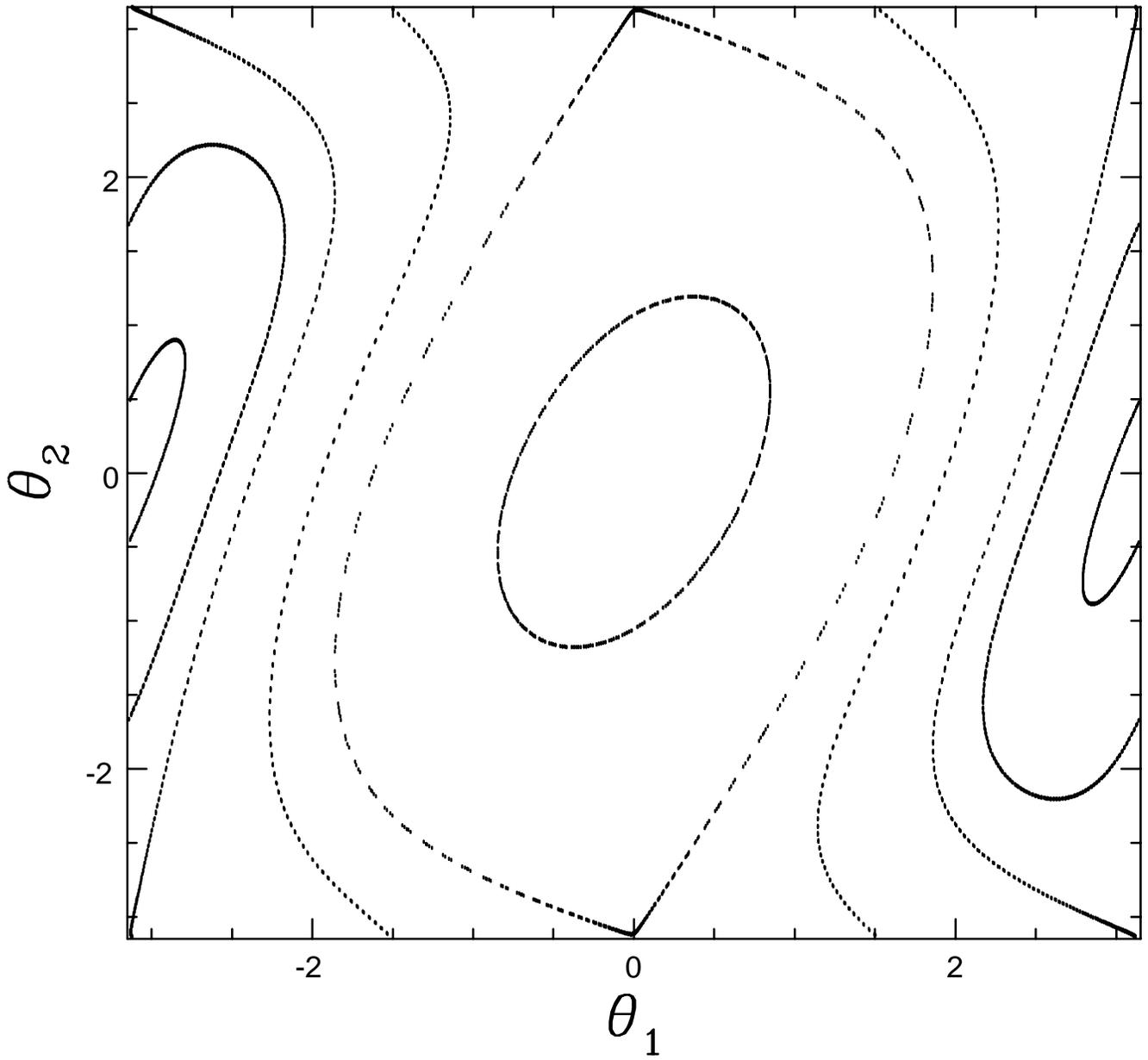
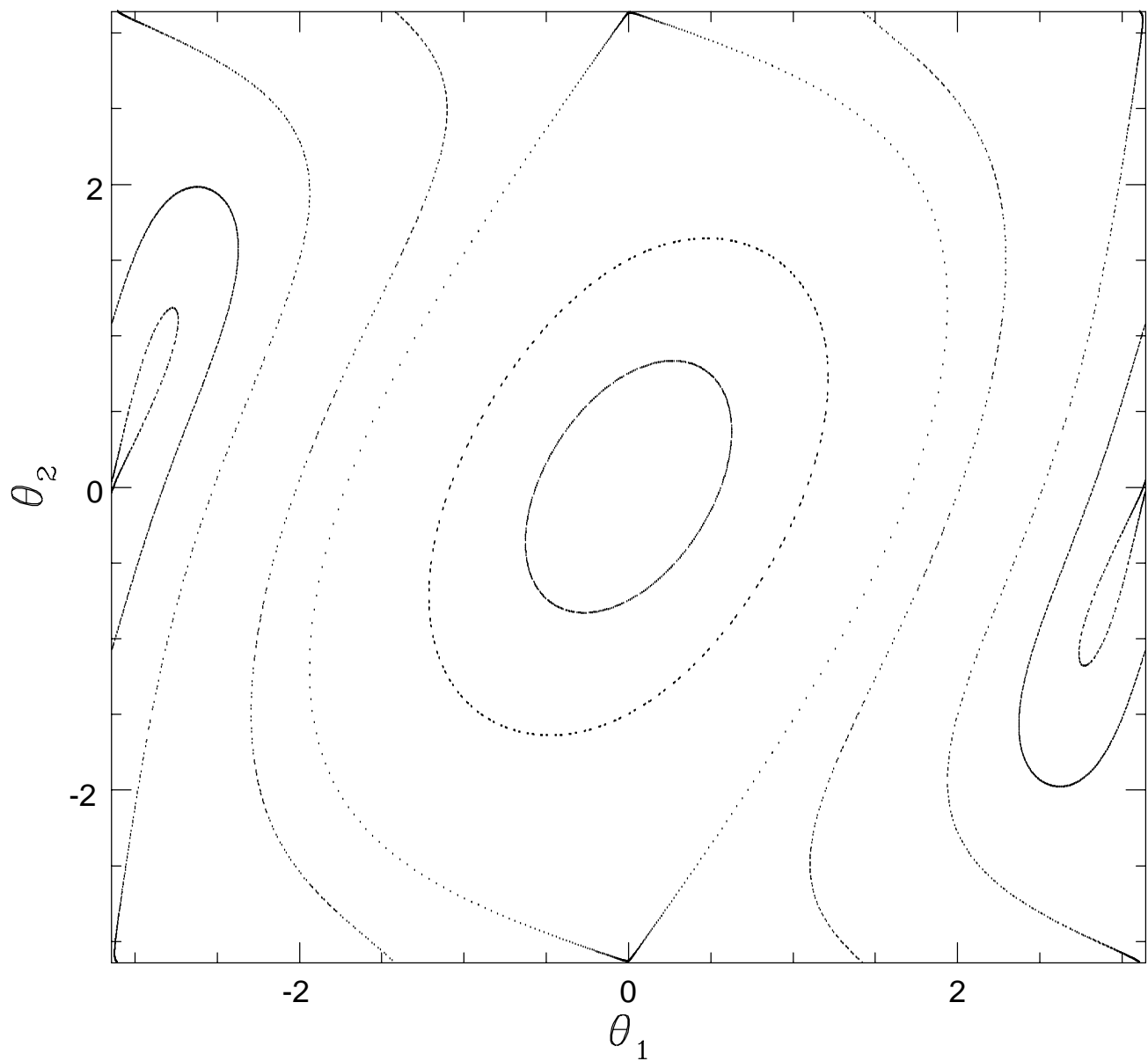
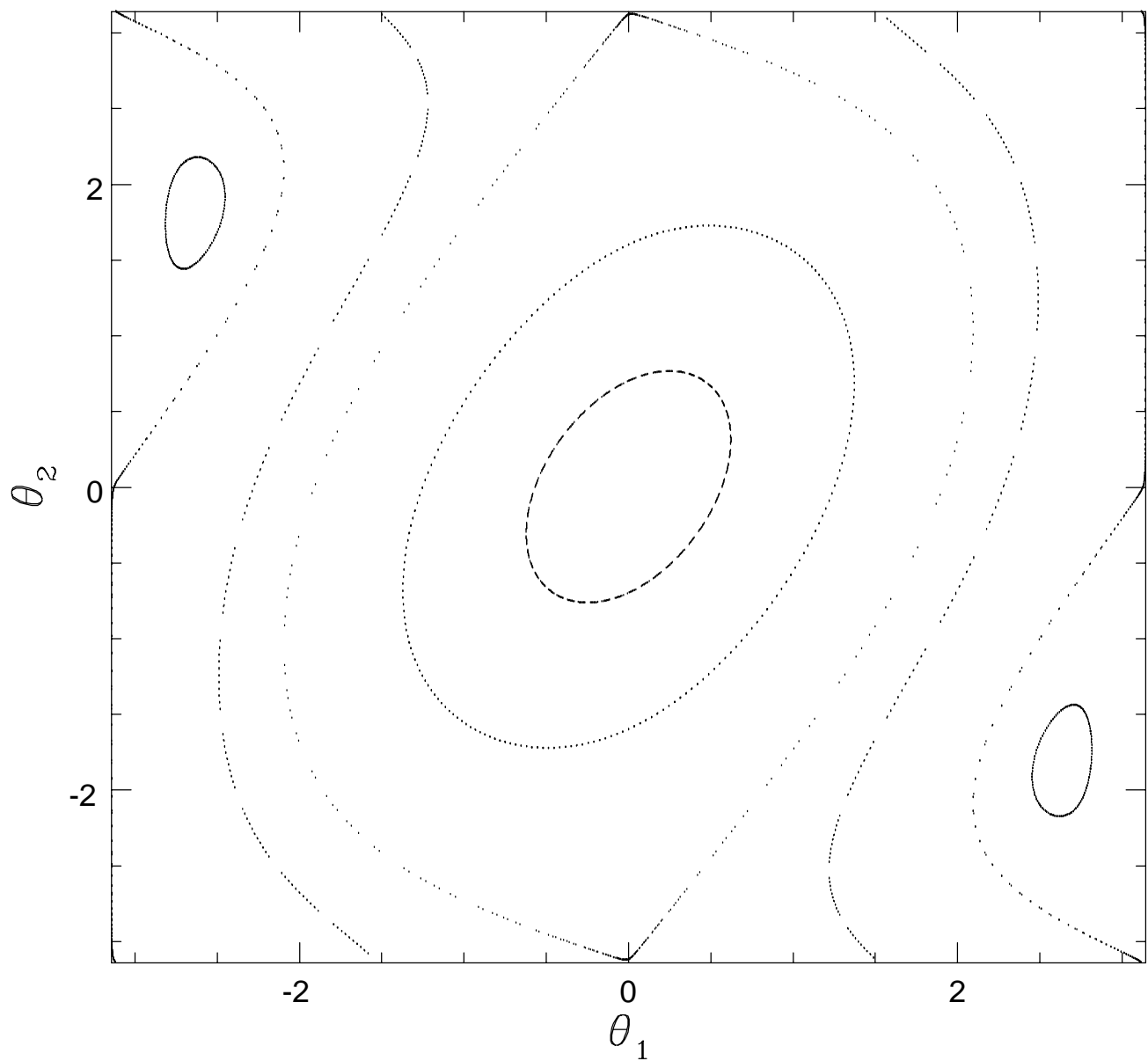
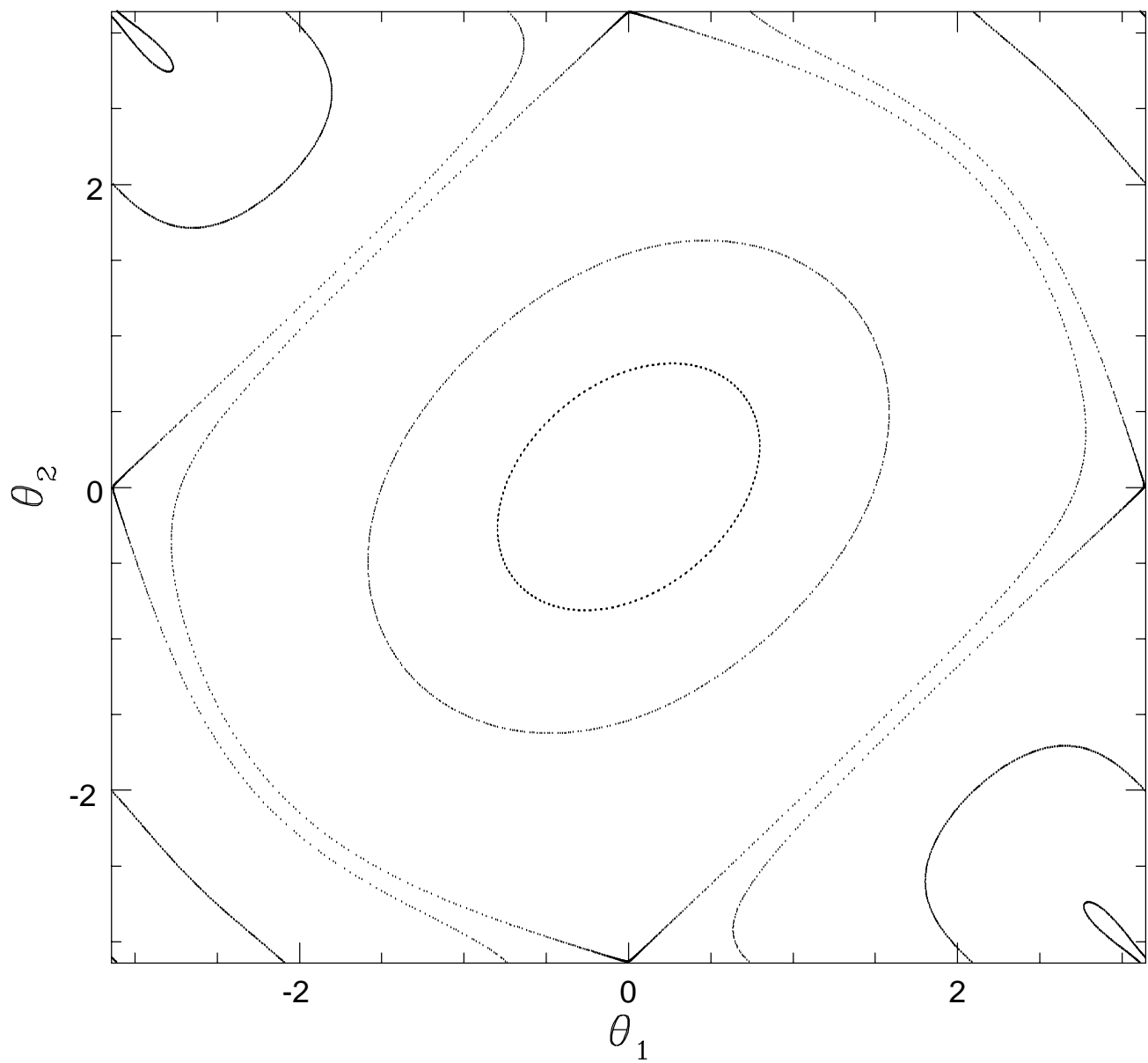
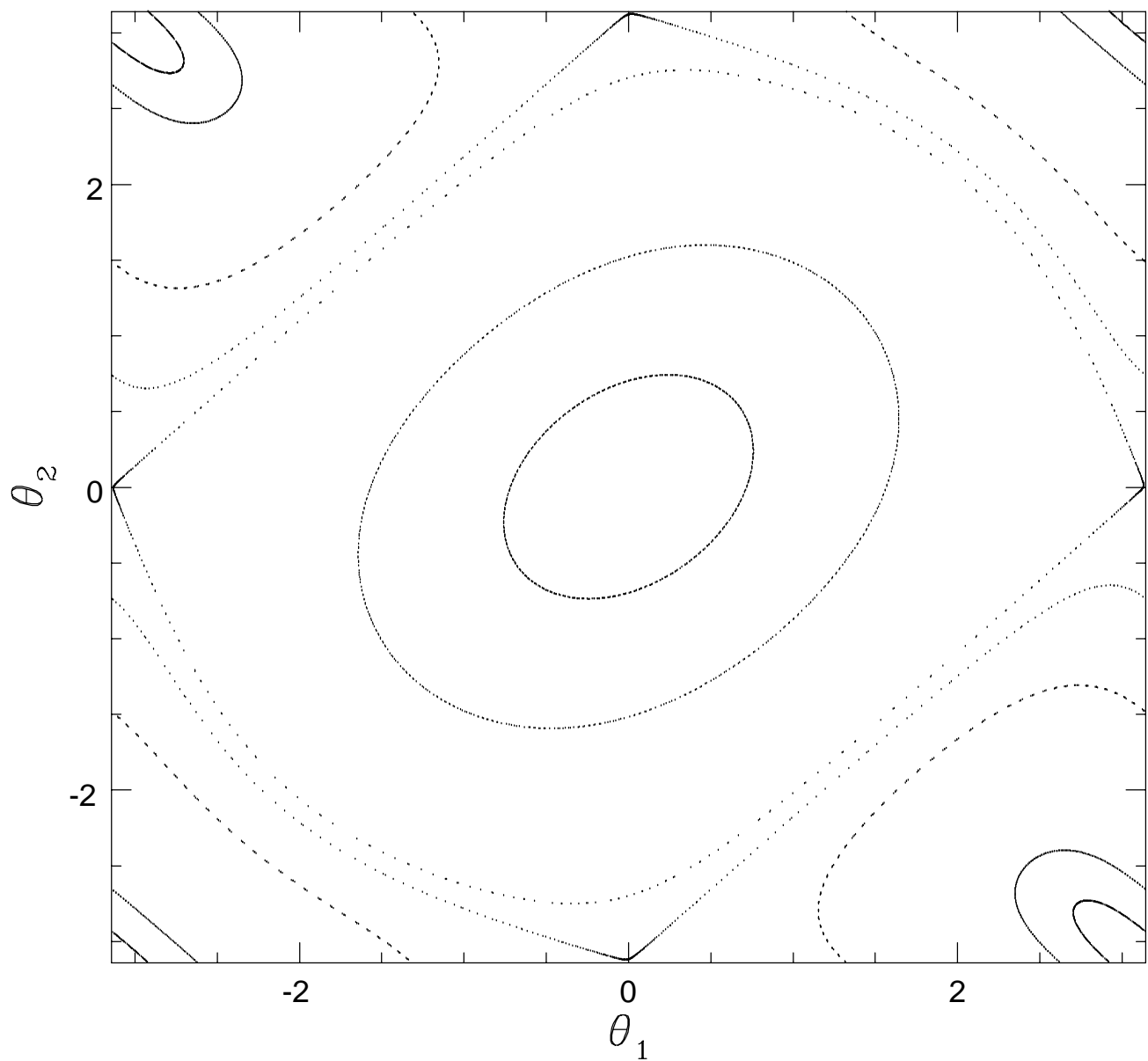
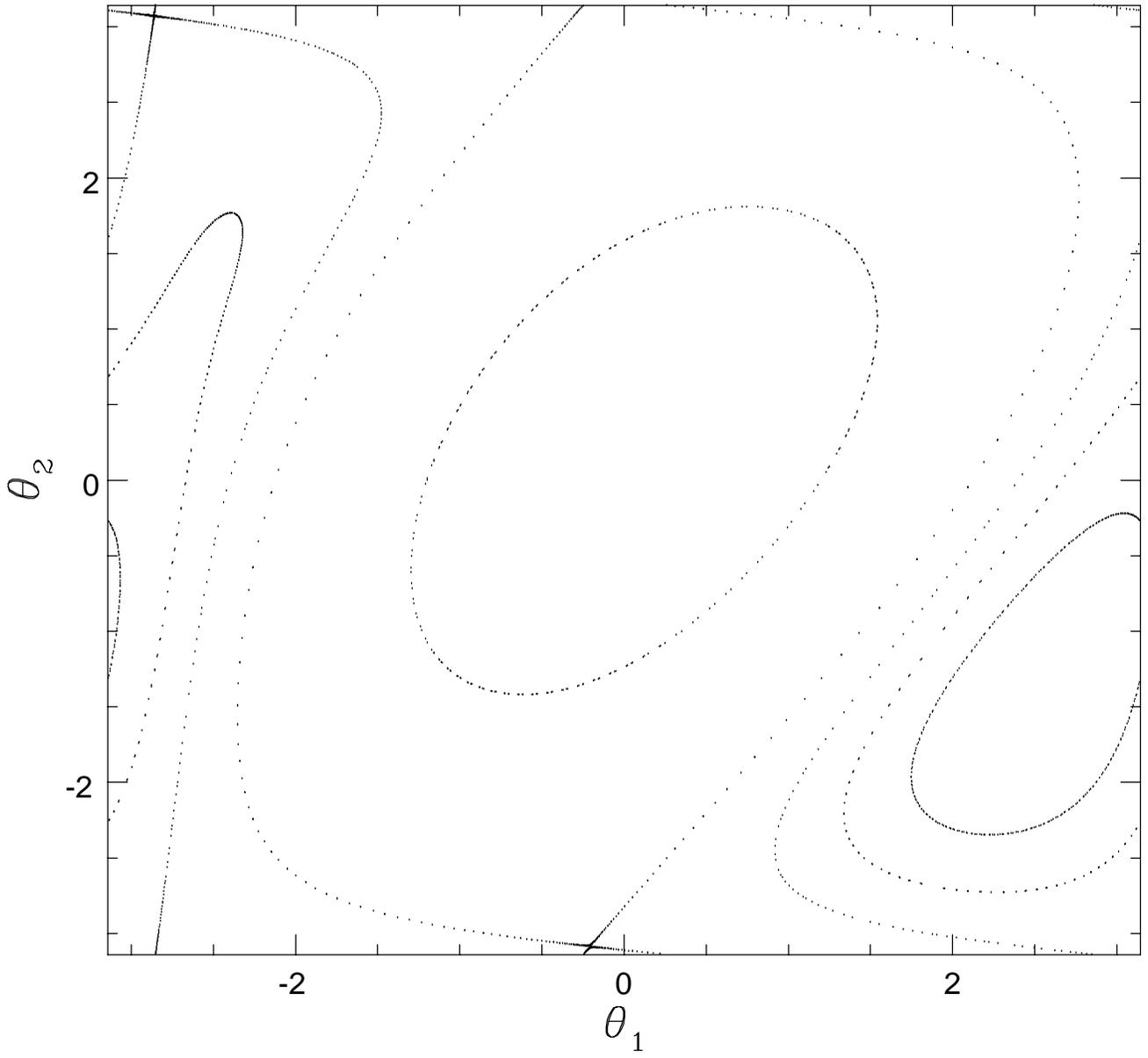
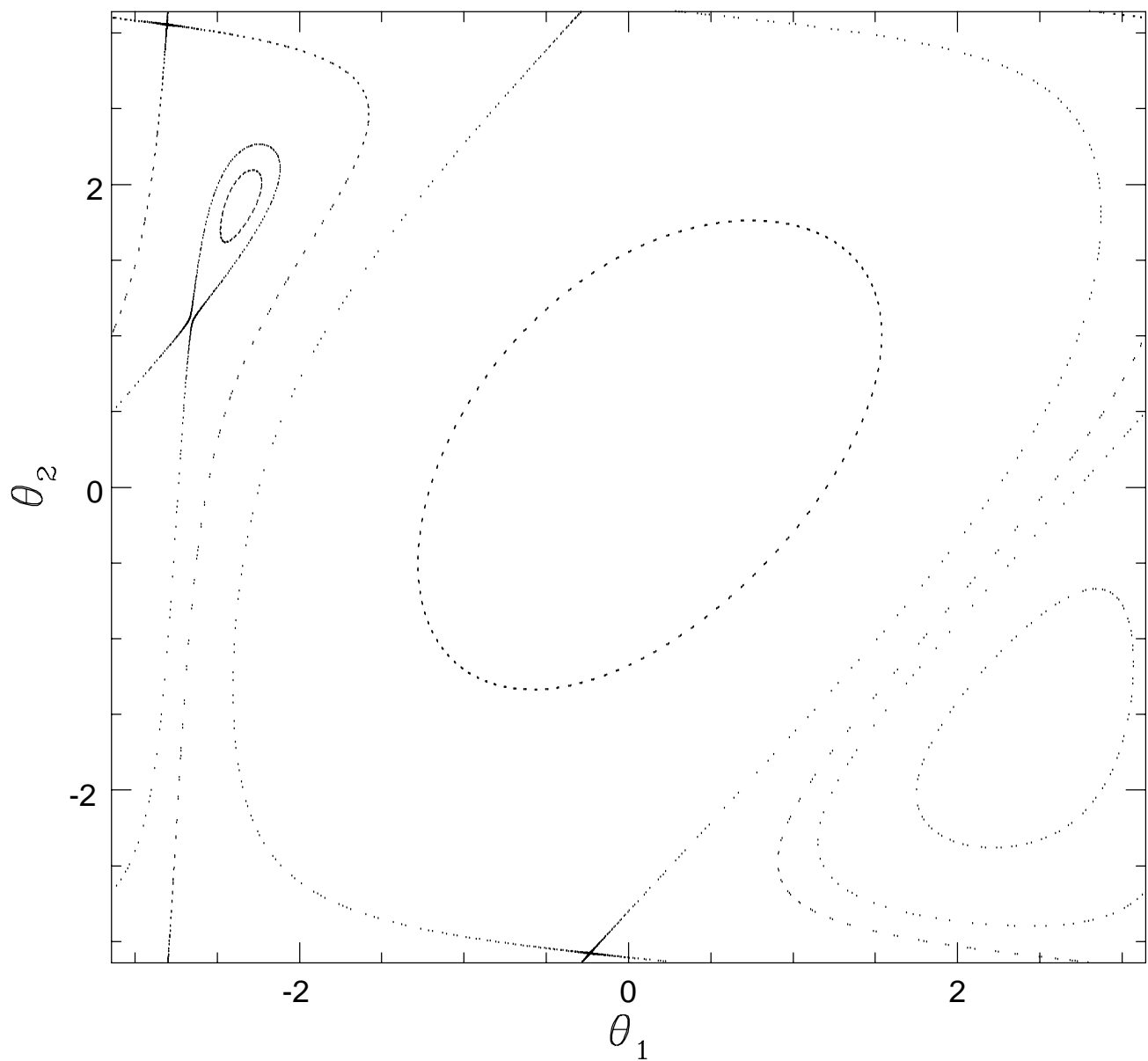
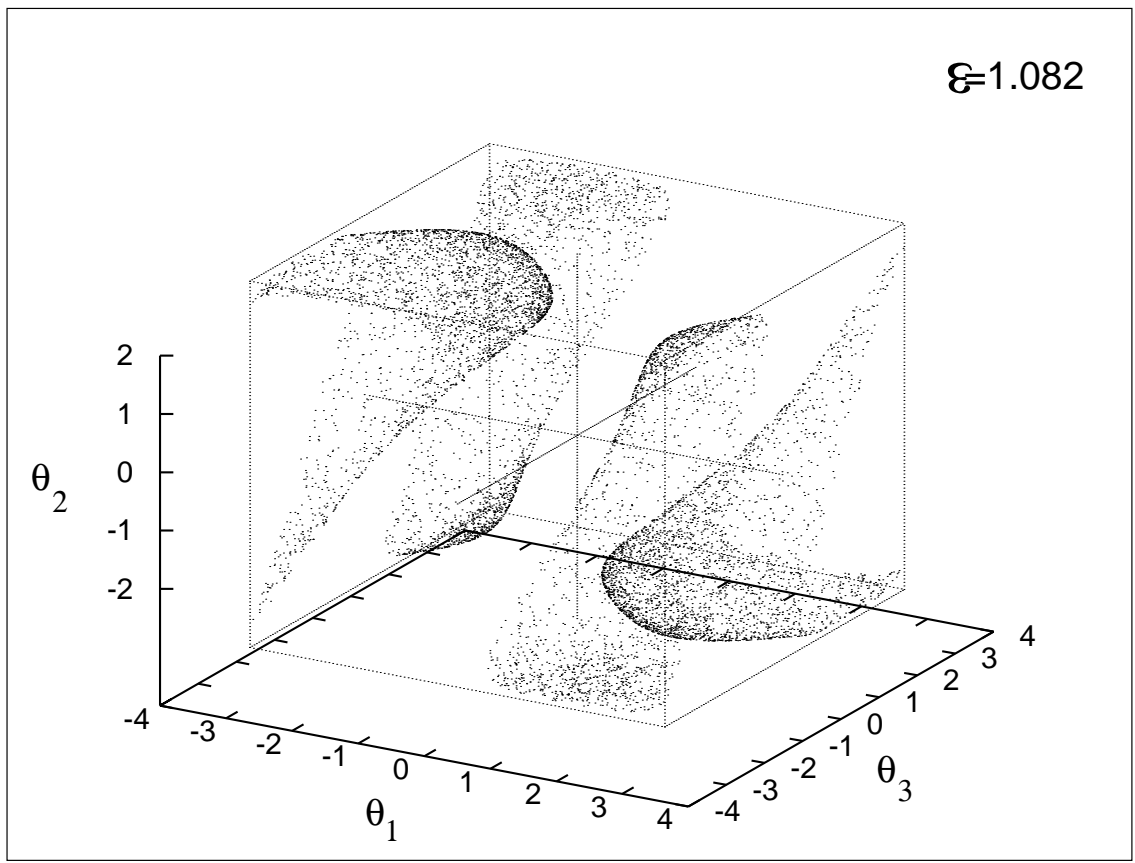
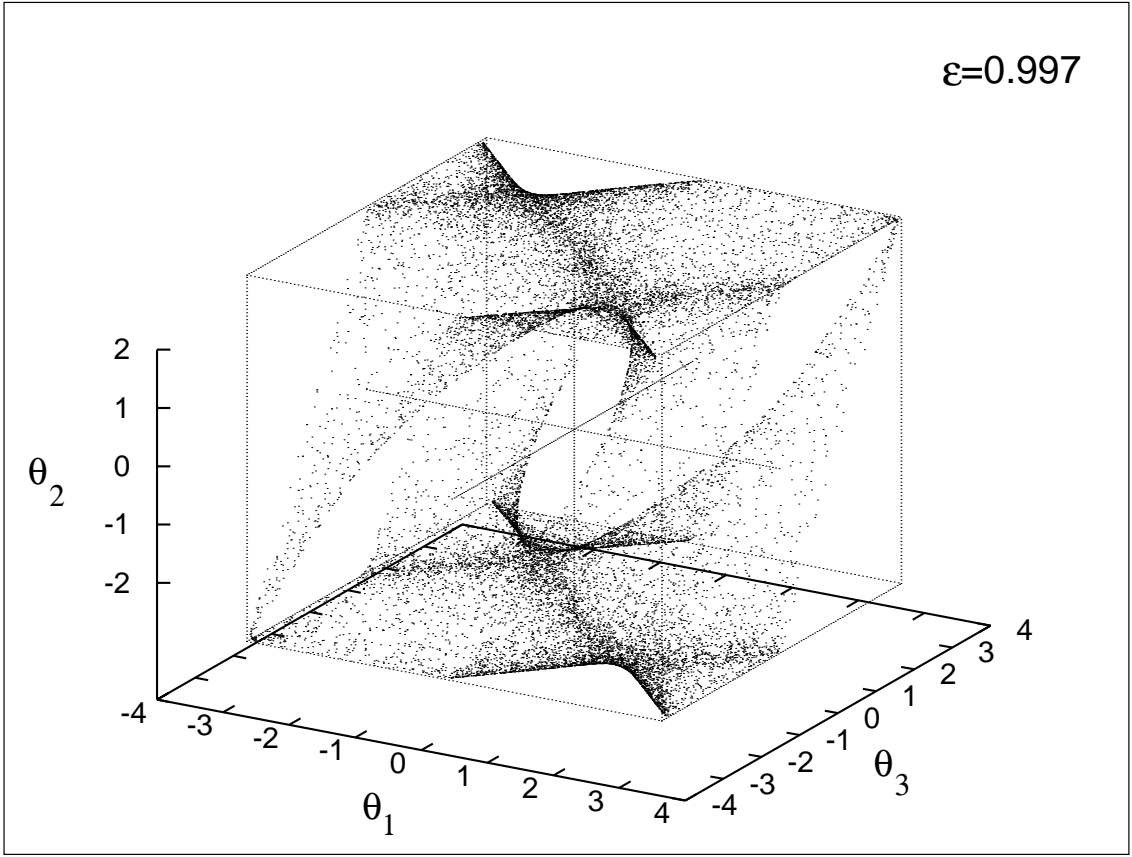
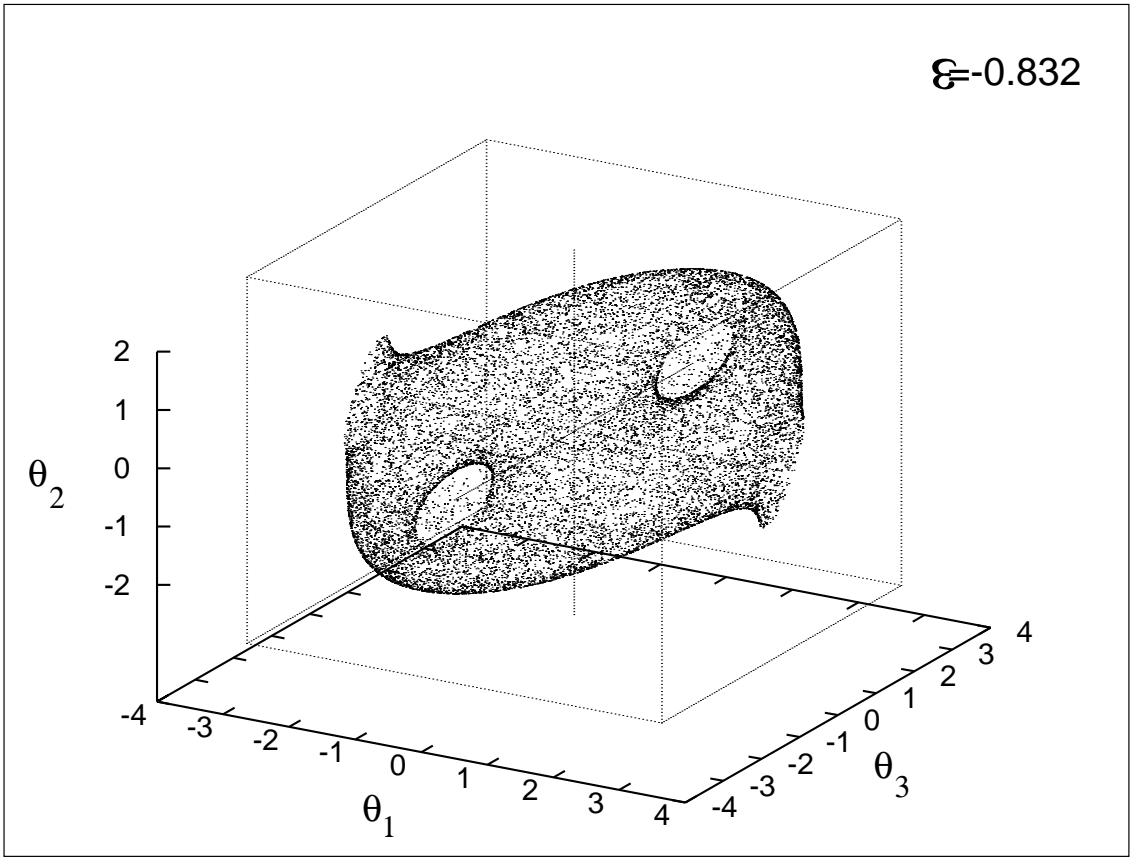
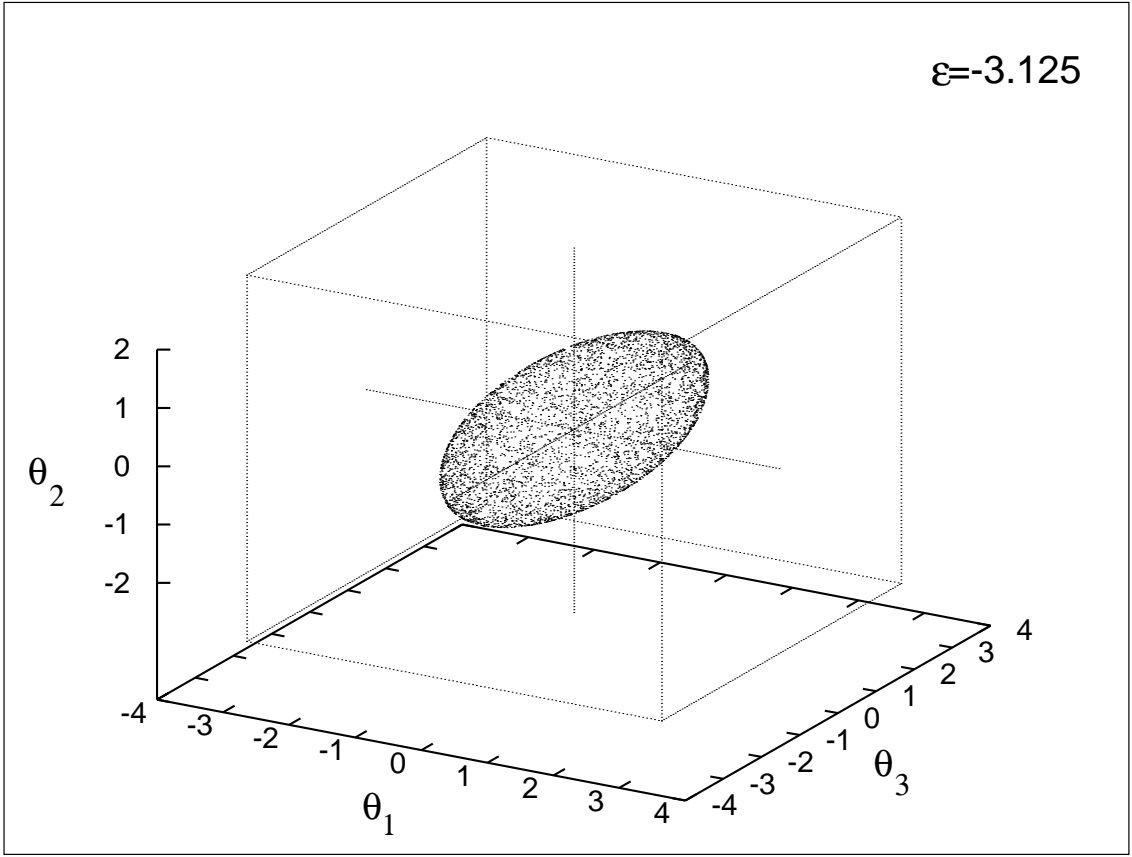
\end{document}